\documentclass{aa}
\usepackage{epsfig,astron,afterpage}

\begin{document}

\thesaurus{ 06 (08.06.2 ; 08 16 05 ; 09.03.1 ; 09.09.1 )}

\title{New Young Stellar Object candidates in the Chamaeleon I molecular cloud 
discovered by  DENIS\thanks{Based on
 observations collected at the European Southern Observatory, La Silla, Chile}} 

\author{L. Cambr\'esy\inst{1} \and E. Copet\inst{1} \and N. Epchtein\inst{2} 
\and B. de Batz\inst{1} \and J. Borsenberger\inst{3} \and P. Fouqu\'e\inst{1,5} 
\and S. Kimeswenger\inst{4} 
\and D. Tiph\`ene\inst{1}}
          
\institute{Observatoire de Paris, F-92195 Meudon Cedex, France \and
Observatoire de la C\^ote d'Azur, BP 4229, F-06304 Nice Cedex, France \and
Institut d'Astrophysique de Paris, 98bis Bd Arago F-75014 Paris, France \and
Institut f\"ur Astronomie der Leopold-Franzens-Universit\"at Innsbruck, 
Technikerstra{\ss}e 25, A-6020 Innsbruck, Austria\and
European Southern Observatory, La Silla, Chile}
\offprints{Laurent Cambr\'esy,\\  Laurent.Cambresy@obspm.fr}
            
\date{Received 19 January 1998}

\titlerunning{New YSO candidates in the Chamaeleon I cloud}
\maketitle

\begin{abstract}

This paper presents an  analysis of the nature of point sources discovered by
DENIS in  an area of $\approx$~$1\fdg5 \times 3\fdg$  around the Chamaeleon I
molecular cloud. Most of the $30\,000$ objects detected in the $J$ band are
background stars that were previously used to derive an accurate extinction
map  of the full area (Cambr\'esy et al., 1997) using star count method.
In the present work, we investigate the young stellar population of the
cloud using the $IJK_{\rm s}$ photometric DENIS data. The whole sample of 126 already
known YSOs, which are mainly T Tauri stars, are identified in the DENIS
catalogue. Besides, we propose 54 sources as new candidates of YSOs.
These new faint objects are selected according to their extremely red near
infrared colour, that cannot be explained only by the reddening of the cloud.
Moreover they are concentrated near to the most obscured areas of the cloud. 
Although
spectroscopic confirmation is badly requested, they are interpreted as probable
classical T Tauri stars that have escaped previous spectroscopic, IRAS or
X-ray observations and pertaining to the low-end of the luminosity function.
Assuming that they are reliable YSOs with massive accretion disks, and using 
theoretical pre-main-sequence  tracks we estimate the age of this sample
to be ranging from $5\, 10^5$ to $4\, 10^6$ years, an elapse of time that 
spans the range between  the end of the star formation in the cloud and
the maximum life time of the circumstellar disk.


\end{abstract}

\section{Introduction}
\begin{figure*}[htb]
	\epsfig{figure=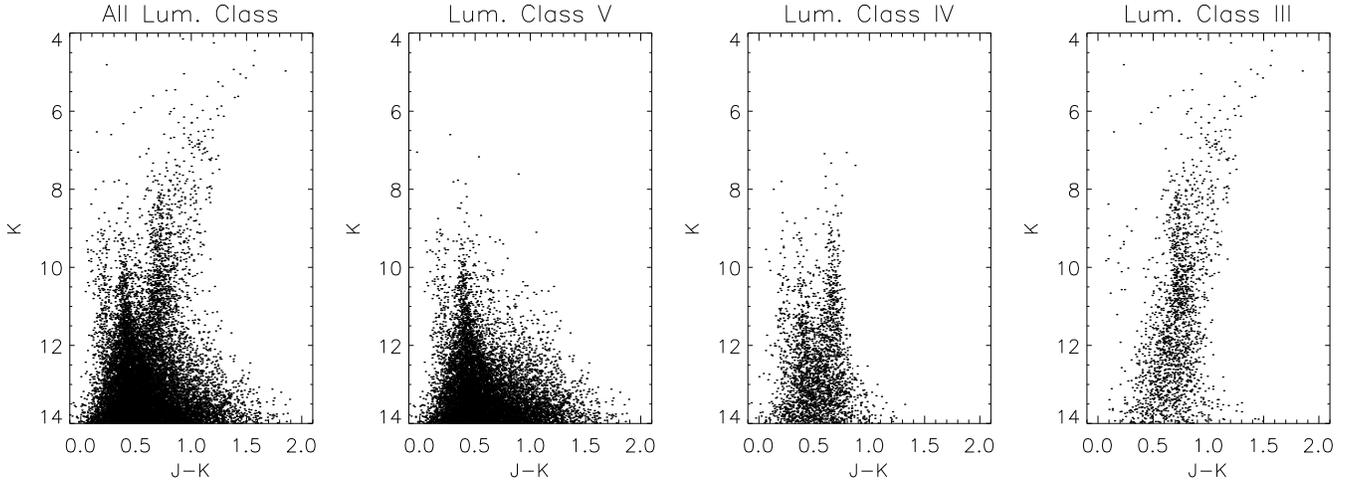,width=18cm}
\caption[]{Colour-magnitude diagram obtained with  the {\em Besan\c{c}on} model
	for the direction of the Chamaeleon I cloud (left), for different 
	luminosity classes (others)}
\label{besancon}
\end{figure*}
\noindent
The Chamaeleon  Molecular Cloud is one of the most  interesting target to 
investigate the 
formation of low mass stars  and the low mass end of the initial mass function 
(IMF), because of its nearby location and its position  far away of the
galactic plane ($b \approx -17 \degr$) where the density of background stars
is relatively small.  A complete census of young objects requires to survey
deeply wide areas in a spectral range that is not much hampered by dust
extinction, namely the near and mid--infrared. This has been made possible 
only recently thanks to the  release of data provided by large scale
sky surveys in the near--infrared  such as DENIS \cite{Epc97}.

Low mass young stellar objects (YSOs) are characterised by ${\rm H}\alpha$
line emission \cite{Har93} and an infrared excess that reveals the presence of
circumstellar material in the near-infrared \cite{WKK+87,WPW91}, or in the mid
and far-infrared as shown by the IRAS \cite{Ryd80,BBW+84,AWPW90,PCW+91} and ISO
\cite{NOA+96} data. X-ray emission  surveys  provided by the  Einstein and
ROSAT missions, have also  contributed to identifying  young objects  
\cite{LFH96,AKC+97}.
Finally, millimetre observations \cite{MLT89,HPZP93}, mostly in CO lines, allow 
the  identification through the measurement of   the gas emission  and outflows 
that often  characterise YSOs.

Known pre-main-sequence stars in the Chamaeleon cloud consist  essentially  of 
T Tauri stars (TTS) \cite{AM89} of  spectral type ranging from K5 to M5 
\cite{AKJ83}. Their mass ranges from 0.2 to $2 M_{\sun}$ with a
distribution peaking at $0.5 M_{\sun}$, and their current age estimate
is greater than $10^6$ years, according to Lawson et al. \cite*{LFH96}.

The Chamaeleon I (Cha I) cloud has been surveyed  in various spectral ranges
from millimetre to X-ray,  but investigation are yet limited to small areas or
low sensitivity.

During its first year of operations, DENIS has covered the whole surface of the 
Cha I cloud in the $IJK_{\rm s}$ bands with a good sensitivity ($K_{\rm s} < 13.5$). Thanks
to its wide surface coverage, stars far away from the known cores of the cloud
can be detected.  An  extinction map of the cloud
was recently  drawn out using $J$ star counts \cite{CEC+97}, and the aim of the 
present work is to pursue the exploitation of these data in order to pick up
already catalogued TTS and to provide an homogeneous set of data and to try to
single out new YSO candidates, especially toward the low luminosity end. 

Up to now, 126 pre-main-sequence stars have been recognised  
in the Cha I cloud by various authors. Section \ref{obs_res} presents the  
DENIS photometric data of these 126 stars and of 54 new candidates that have
been selected using the  method  described in Section \ref{selection}.
The nature of the sources is discussed in Section \ref{nature} and some
constraints on the circumstellar environments are derived in Sect. 
\ref{circum}. Finally,  in Sect. \ref{KLF}, we discuss the luminosity
function, and we  estimate the age of the period of star formation in the cloud.

\section{Observations and Results}
\label{obs_res}

The observations presented here have been collected as part of the DENIS survey 
between January and May 1996 at La Silla (Chile) using the ESO 1 metre 
telescope equipped with the specially designed 3-channel camera
\cite{Epc97,CER+98}. They cover an area of $1\fdg45 \times 2\fdg94$ centred
at $\alpha=11^{\rm h}06^{\rm m} \, ,\, \delta=-77\degr 30^{\rm m}$
(J2000) in three bands, namely $I$ (0.8$\ \mu {\rm m}$), $J$ (1.25$\ \mu {\rm m}$) and
$K_{\rm s}$ (2.15$\ \mu {\rm m}$).
They consist of 13 strips each involving 180 images of $12'\times12'$ taken
at constant RA along an arc of $30\degr$ in declination. The overlap 
between two adjacent strips reaches $75 \%
$ in the Cha I cloud because of the proximity of the south pole. Wherever a
star is picked up in two adjacent strips, position and flux values are 
averaged. Limiting magnitudes are 18, 16 and 13.5 at $3\sigma$ in $I$, $J$
and $K_{\rm s}$ bands, respectively, and a conservative estimate of the 
position accuracy is $1\arcsec$  {\it rms} in both directions. The numbers
of stars detected in $IJK_{\rm s}$ are $\sim 50\,000$, $30\,000$ and $10\,000$, 
respectively.
Table \ref{knownYSO} displays a list of known stars associated with the Cha I
cloud with the DENIS positions and $IJK_{\rm s}$ photometry. Names given in column 4
refer to previous investigations quoted in the footnote 1 of the Table. Column 9
lists the cloud extinction as measured on our map \cite{CEC+97}, and col. 10
the extinction derived from the $I-J$ colour excess.
Table \ref{newYSO} presents the list of our new candidates, their DENIS
positions and photometry, and the extinction derived in the same way as in
Table \ref{knownYSO}.\\

\begin{table*}
	\caption[]{List of known T Tauri stars}
	\label{knownYSO}
%%%%%%%%%%%%%%%%%%%%%%%%%%%%%%%%%%%%%%%%%%%%%%%%%%%%%%%%%%%%%%%%%%%%%%%%%%%%%%%%
%%%%% Debut de la table 1 (known TTS) %%%%%%%%%%%%%%%%%%%%%%%%%%%%%%%%%%%%%%%%%%
%%%%%%%%%%%%%%%%%%%%%%%%%%%%%%%%%%%%%%%%%%%%%%%%%%%%%%%%%%%%%%%%%%%%%%%%%%%%%%%%
\begin{tabular}{lcclrrrrrr}
\hline
Nb & R.A. (J2000) & Dec. (J2000) & Name\footnotemark & $V$ & $I$ &$J$ &$K_{\rm s}$ & 
Av$_{cl}$ &Av$_*$\\
(1)&(2)&(3) &(4) &(5) &(6) &(7) &(8) &(9) & (10)\\
\hline 
1\footnotemark    & 10 52 03.0  & -77 09 49  &  \object{T1}, \object{Sz1}  &  &  &  &  &  & \\ 
2\footnotemark    & 10 54 30.0  & -77 55 18  &  \object{T2}, \object{SW Cha}  &  &  &  &  &  0.0  & \\ 
3\footnotemark    & 10 56 01.1  & -77 24 38  &  \object{T3}, \object{SX Cha}  &  14.65 &  & 10.20 & 8.38 &  0.9 & \\ 
4\footnotemark    & 10 56 32.0  & -77 11 38  &  \object{T4}, \object{SY Cha}  &  13.03 &  & 10.09 & 8.74 &  2.7 & \\
5    & 10 57 42.7  & -76 59 36  &  \object{T5}, \object{Sz4}  &  &  12.18  &  10.31  &  9.04  &  1.2  & 3.3  \\
6  & 10 58 05.9  & -77 28 24  &  \object{CHRX3}  &  12.26  &  9.99  & 8.33  & 7.04  & 1.4  & 2.6\\
7    & 10 58 17.1  & -77 17 17  &  \object{T6}, \object{SZ Cha}  &  11.99  &  10.32 &  8.96 &  7.43 &  1.0  & 1.7 \\
8    & 10 59 01.4  & -77 22 41  &  \object{T7}, \object{TW Cha}  &  13.08  &  11.26  & 9.87  & 8.18  & 1.4  & 1.8\\
9    & 10 59 07.5  & -77 01 40  &  \object{T8}, \object{Sz6}, \object{CHX3}  &  11.22  &  9.82  & 8.40  & 7.16  & 0.9  & 1.9\\
10    & 11 00 14.5  & -76 44 15  &  \object{T9}, \object{Sz7}  &  &  12.22  & 8.96  & 7.38  & 0.5  & 7.4\\
11  & 11 00 14.9  & -77 14 38  &  \object{CHXR8}  &  11.45  &  10.49  &  9.85   & 9.45  &  1.4  & 0.0\\
12   & 11 00 40.7  & -76 19 28  &  \object{T10}, \object{Sz8}  &  &  13.47  & 11.82  & 10.69  & 0.1  & 2.5\\
13  & 11 01 19.0  & -76 27 03  &  \object{CHXR9C}  &  14.11  &  11.60  & 10.04  & 8.81  & 0.5  & 2.3\\
14  & 11 02 15.4  & -77 10 59  &  \object{B9}  &  &  14.80  &  13.51  &  12.62  &  0.2  & 1.5\\
15   & 11 02 25.4  & -77 33 36  &  \object{T11}, \object{CS Cha}  &  11.63  &  10.12  & 9.00  & 8.26  & 5.0 & 1.0 \\
16   & 11 02 33.0  & -77 29 13  &  \object{Hn1}  &  &  12.97  & 11.23  & 10.16  & 4.3  & 3.0\\
17   & 11 02 55.5  & -77 21 51  &  \object{T12}, \object{Sz10}  &  &  13.14  & 11.41  & 10.36  & 1.4  & 3.0\\
18  & 11 03 12.0  & -77 21 05  &  \object{CHXR11}  &  11.52  &  9.58  &  8.11  &  7.15  &  1.3  & 2.0\\
19   & 11 03 48.1  & -77 19 57  &  \object{Hn2}  &  &  13.91  & 11.26  & 9.96  & 1.4  & 5.5\\
20\footnotemark   & 11 03 51.4  & -76 55 46  &  \object{T13}, \object{TZ Cha}  &  &  15.04  & 14.35  &   & 0.1  &  0.0\\
21   & 11 03 57.3  & -77 21 34  &  \object{Hn3}  &  &  12.51  & 10.75  & 9.69  & 1.9  & 2.9\\
22   & 11 04 09.5  & -76 27 19  &  \object{T14}, \object{CT Cha}  &  12.36  &  10.89  & 9.63  & 8.54  & 0.1  &  1.4\\
23  & 11 04 11.6  & -76 54 32  &  \object{CHXR72}  &  15.04  &  13.09  &  11.79  & 10.96  &  0.0  &  1.2 \\
24   & 11 04 23.2  & -77 18 08  &  \object{T14a}, \object{HH48}  &  &  16.25  &  14.48  &  12.36  &  2.2  & 2.9\\
25   & 11 04 24.7  & -77 25 49  &  \object{T15}, \object{Sz12}  &  &  13.35  & 10.87  & 9.45  & 3.8  & 5.0\\
26  & 11 04 43.0  & -77 41 57  &  \object{B18}  &  &  13.63  &  11.78  &  10.61  &  6.1  & 3.1\\
27  & 11 04 51.5  & -76 25 24  &  \object{CHXR14N}  &  14.13  &  11.89  & 10.59  & 9.61  & 0.5  & 1.5\\
28  & 11 04 53.2  & -76 25 51  &  \object{CHXR14S}  &  14.66  &  12.05  & 10.67  & 9.64  & 0.5  & 1.7\\
29   & 11 04 57.4  & -77 15 57  &  \object{T16}, \object{Sz13}  &  &  14.80  & 12.10  & 10.27  & 2.9  & 5.7\\
30   & 11 05 15.1  & -77 11 29  &  \object{Hn4}  &  &  13.38  & 10.93  & 9.46  & 2.8  & 4.9\\
31   & 11 05 15.6  & -77 52 55  &  \object{T18}, \object{Sz14}  &  &  15.54  & 13.01  & 11.25  & 4.4  & 5.2\\
32   & 11 05 22.0  & -76 30 22  &  \object{T17}, \object{UU Cha}  &  &  14.36  & 13.44  & 12.61  & 1.4  & 0.4\\
33   & 11 05 41.9  & -77 54 44  &  \object{T19}, \object{Sz15}  &  &  12.35  & 11.29  & 10.64  & 4.3  & 0.8\\
34  & 11 05 43.4  & -77 26 52  &  \object{CHXR15}  &  17.27  &  13.53  & 11.29  & 10.09  & 6.0  & 4.2\\
35   & 11 05 53.0  & -76 18 26  &  \object{T20}, \object{UV Cha}  &  13.79  &  11.93  & 10.43  & 9.21  & 0.5  & 2.1\\
36   & 11 06 15.7  & -77 21 57  &  \object{T21}, \object{CHX7}  &  11.28  &  9.42  & 7.66  & 6.27  & 6.2  & 2.9\\
37   & 11 06 42.3  & -76 35 49  &  \object{Hn5}  &  &  14.15  & 11.88  & 10.14  & 2.4  & 4.2\\
38   & 11 06 43.9  & -77 26 35  &  \object{T22}, \object{UX Cha}  &  &  13.02  & 10.94  & 9.27  & 5.8  & 4.0\\
39  & 11 06 45.5  & -77 27 03  &  \object{CHXR20}  &  14.76  &  12.00  &  10.31  &  8.76  &  5.0  & 2.7\\
40  & 11 06 47.1  & -77 22 31  &  \object{Ced110 IRS4}  &  &  &  &  12.95  &  6.9  & \\  
41   & 11 06 59.5  & -77 18 54  &  \object{T23}, \object{UY Cha}  &  &  12.92  & 11.30  & 9.89  & 9.0  & 2.7\\
42  & 11 07 09.8  & -77 23 05  &  \object{Ced110 IRS6}  &  &  &  &  10.94  &  7.1  & \\  
43  & 11 07 10.8  & -77 43 44  &  \object{CHXR22W}  &  &  16.72  &  14.15  &  12.52  &  4.8  & 5.4\\
44   & 11 07 11.8  & -77 46 39  &  \object{Hn6}  &  &  13.34  & 11.17  & 9.59  & 4.5  & 4.0\\
45   & 11 07 12.5  & -76 32 23  &  \object{T24}, \object{UZ Cha}  &  14.90  &  12.49  & 10.86  & 9.23  & 2.8  & 2.4 \\
46  & 11 07 13.7  & -77 43 50  &  \object{CHXR22E}  &  &  14.84  &  11.93  &  9.96  &  4.8  & 6.3\\
47   & 11 07 19.6  & -76 03 05  &  \object{T25}, \object{Sz18}  &  15.35  &  12.61  & 11.04  & 9.84  & 0.6  & 2.4\\
48   & 11 07 21.1  & -77 38 08  &  \object{T26}, \object{Sz19}, \object{CHX9}  &  10.68  &  9.31  & 8.02  & 6.34  & 5.8  & 1.2\\
49   & 11 07 28.6  & -76 52 12  &  \object{T27}, \object{VV Cha}  &  14.80  &  12.26  & 10.73  & 9.60  & 1.4  & 2.5\\
50  & 11 07 33.4  & -77 28 28  &  \object{CHXR25}  &  15.99  &  13.14  &  11.78  &  10.88  &  5.5  & 1.7\\
51  & 11 07 35.7  & -77 34 50  &  \object{CHXR76}  &         &  14.43  &  12.14  & 10.90  &  6.3  &  4.5 \\
52  & 11 07 37.4  & -77 33 34  &  \object{CHXR26}  &  &  15.16  &  11.56  &  9.31  &  6.2  & 8.4\\
53   & 11 07 44.1  & -77 39 42  &  \object{T28}, \object{Sz21}  &  15.34  &  12.56  & 10.20  & 8.34  & 5.0  & 4.6\\
54   & 11 07 56.3  & -77 27 26  &  \object{CHX10a}, \object{F29}  &  &  11.01  &  9.12  &  7.80  &  5.4  & 3.2\\
55   & 11 07 57.7  & -77 17 27  &  \object{Baud38}  &  &  17.42  & 13.08  & 9.87  & 6.5  & 10.6\\
56   & 11 07 58.4  & -77 38 45  &  \object{T29}, \object{Sz22}  &  14.2  &  12.76  & 10.18  & 7.19  & 4.2  &  5.3\\
57   & 11 07 58.5  & -77 42 42  &  \object{T30}, \object{Sz23}, \object{CHX11}  &  &  14.58  & 11.98  & 9.92  & 4.0  &  5.4\\
58  & 11 08 00.4  & -77 17 31  &  \object{CHXR30}  &  &  15.57  &  11.74  &  9.04  &  7.4  & 9.1\\
\hline
\end{tabular}
\end{table*}

\begin{table*}
{\bf Table \ref{knownYSO}.} (continued)\\
\begin{tabular}{lcclrrrrrr}
\hline
Nb & R.A. (J2000) & Dec. (J2000) & Name\footnotemark[1] & $V$ & $I$ &$J$ &$K_{\rm s}$ & 
Av$_{cl}$ & Av$_*$\\
(1)&(2)&(3) &(4) &(5) &(6) &(7) &(8) &(9) & (10)\\
\hline
59   & 11 08 01.9  & -77 42 29  &  \object{T31}, \object{VW Cha}  &  12.51  &  10.56  & 8.66  & 6.98  & 4.0  &  9.1\\
60   & 11 08 03.7  & -77 39 18  &  \object{T32}, \object{HD 97048}  &  8.45  &  8.58  & 7.39  & 6.26  & 4.2  & 0.7\\
61   & 11 08 15.8  & -77 33 54  &  \object{T33}, \object{CHX12}  &  &  10.54  & 8.58  & 6.22  & 5.4  & 3.3\\
62   & 11 08 17.0  & -77 44 37  &  \object{T34}, \object{Sz26}  &  &  13.04  & 11.03  & 10.05  & 4.0  & 3.6\\
63   & 11 08 37.1  & -77 43 51  &  \object{IRN}  &  &  15.86  & 11.93  & 8.48  & 3.9  & 9.4 \\
64   & 11 08 39.5  & -77 16 05  &  \object{T35}, \object{Sz27}  &  16.28  &  13.22  & 11.89  & 8.89  & 4.7  & 4.9\\
65   & 11 08 41.1  & -76 36 08  &  \object{CHX13a}  &  15.56  &  12.54  & 10.45  & 9.09  & 3.3  & 3.2\\
66\footnotemark   & 11 08 49.7  & -76 44 28  &  \object{T36}, \object{VX Cha}  &  &  &  &  &  2.0  & \\
67\footnotemark   & 11 08 51.2  & -77 43 39  &  \object{C9-1}  &  &  &  &  &  3.6  & \\
68   & 11 08 51.3  & -76 25 14  &  \object{T37}, \object{Sz28}  &  &  14.74  & 12.45  & 11.30  & 1.1  & 4.3 \\
69  & 11 08 54.7  & -77 32 12  &  \object{CHXR78C}  &  &  14.75  &  12.17  &  10.92  &  4.0  & 4.9\\
70   & 11 08 55.1  & -77 02 13  &  \object{T38}, \object{VY Cha}  &  &  12.99  & 11.12  & 9.26  & 2.3  & 3.0\\
71   & 11 09 05.5  & -77 09 58  &  \object{Hn7}  &  &  13.83  & 11.79  & 10.77  & 2.6  & 3.5\\
72   & 11 09 12.3  & -77 29 13  &  \object{T39}, \object{Sz30}, \object{CHX14}  &  13.17  &  11.22  & 9.72  & 8.75  & 3.9  & 1.8\\
73   & 11 09 14.2  & -76 28 40  &  \object{Hn8}  &  &  13.69  & 11.64  & 10.67  & 2.3  & 3.4\\
74  & 11 09 18.1  & -76 27 58  &  \object{CHXR37}  &  13.92  &  11.66  &  9.87  & 8.41  & 2.3  & 2.7\\
75   & 11 09 18.6  & -76 30 29  &  \object{Hn9}  &  &  14.91  & 11.63  & 8.95  & 4.9  & 7.3\\
76   & 11 09 23.0  & -76 34 32  &  \object{C1-6}  &  &  17.99  & 13.13  & 9.32  & 4.0  & 11.4\\
77   & 11 09 24.2  & -76 23 21  &  \object{T40}, \object{VZ Cha}  &  12.75  &  11.43  & 9.82  & 7.71  & 0.8  & 2.4\\
78  & 11 09 40.5  & -76 28 39  &  \object{CHXR40}  &  14.04  &  11.47  &  9.95  & 8.73  & 2.3  & 2.0\\
79   & 11 09 42.4  & -76 34 58  &  \object{C1-25}  &  &   & 13.49  & 9.80  & 5.6  & \\
80   & 11 09 43.0  & -77 25 59  &  \object{C7-1}  &  &  15.71  & 12.19  & 10.42  & 5.1  & 7.8\\
81   & 11 09 46.2  & -76 43 54  &  \object{C2-3}  &  &  14.50  & 11.75  & 9.96  & 2.9  & 5.7\\
82   & 11 09 46.6  & -76 34 46  &  \object{Hn10E}  &  &  14.81  & 11.93  & 9.84  & 5.6  & 6.3\\
83\footnotemark   & 11 09 46.6  & -76 34 46  &  \object{Hn10W}  &  &  &  &  &  5.6  & \\
84   & 11 09 47.8  & -77 26 30  &  \object{Baud43}  &  &  16.02  & 12.38  & 9.93  & 4.5  & 7.9 \\
85   & 11 09 50.4  & -76 36 48  &  \object{T41}, \object{HD 97300}  &  &  8.84  & 7.56  & 7.05  & 4.1  & 1.4\\
86   & 11 09 53.8  & -76 34 25  &  \object{T42}, \object{Sz32}  &  &  14.61  & 10.62  & 6.42  & 5.8  & 9.6\\
87   & 11 09 54.4  & -76 29 25  &  \object{T43}, \object{Sz33}, \object{CHX15}  &  &  13.99  & 11.29  & 9.17  & 2.8  & 5.7\\
88   & 11 09 55.5  & -76 32 41  &  \object{C1-2}  &  &   & 13.87  & 9.52  & 6.1  & \\
89   & 11 09 59.7  & -77 37 09  &  \object{T45}, \object{WX Cha}  &  14.86  &  11.82  & 9.69  & 7.86  & 3.5  & 3.2\\
90   & 11 10 01.0  & -76 34 58  &  \object{T44}, \object{WW Cha}  &  13.27  &  11.02  & 8.56  & 5.75  & 5.8  & 5.0\\
91   & 11 10 04.1  & -76 33 29  &  \object{Hn11}  &  &  14.46  &  11.55  &  9.39  &  6.2  & 6.3\\
92   & 11 10 05.1  & -76 35 45  &  \object{T45a}  &  14.34  &  11.93  & 10.23  & 9.04  & 5.4  & 2.7\\
93   & 11 10 07.4  & -76 29 38  &  \object{T46}, \object{WY Cha}  &  13.98  &  11.60  & 9.87  & 8.31  & 2.8  & 2.8\\
94   & 11 10 28.9  & -77 17 00  &  \object{Hn12W}  &  &  13.85  & 11.60  & 10.64  & 5.4  & 4.3\\
95\footnotemark   & 11 10 28.9  & -77 17 00  &  \object{Hn12E}  &  &  &  &  &  5.4  & \\
96  & 11 10 38.4  & -77 32 40  &  \object{CHXR47}, \object{F34}  &  14.42  &  11.66  & 9.59  & 8.23  & 2.8  & 3.8\\
97   & 11 10 50.0  & -77 17 53  &  \object{T47}, \object{Sz37}  &  15.54  &  13.01  & 10.91  & 9.07  & 5.0  & 3.8\\
98   & 11 10 53.7  & -76 34 32  &  \object{T48}, \object{WZ Cha}  &  15.26  &  12.76  & 11.08  & 9.95  & 3.5  & 2.6\\
99   & 11 10 56.4  & -76 45 33  &  \object{Hn13}  &  &  13.85  & 11.13  & 9.96  & 4.2  & 5.8\\
100   & 11 11 14.6  & -76 41 11  &  \object{Hn14}  &  &  16.25  & 14.17  & 12.70  & 3.4  & 3.8\\
101  & 11 11 35.2  & -76 36 21  &  \object{CHXR48}  &  14.73  &  12.35  & 10.79  & 9.65  & 2.0  & 2.3\\
102   & 11 11 40.1  & -76 20 15  &  \object{T49}, \object{XX Cha}  &  15.28  &  12.56  & 10.68  & 9.24  & 0.2  & 3.2\\
103  & 11 11 46.8  & -76 20 09  &  \object{CHX18N}  &  12.05  &  10.37  & 9.02  & 7.75  & 0.4  & 1.6\\
104   & 11 11 54.5  & -76 19 31  &  \object{Hn15}  &  14.41  &  11.88  & 10.14  & 9.12  & 0.4  & 2.8\\
105   & 11 12 03.7  & -76 37 03  &  \object{Hn16}  &  &  13.97  & 11.69  & 10.66  & 1.8  & 4.4\\
106   & 11 12 10.3  & -76 34 37  &  \object{T50}, \object{Sz40}  &  &  13.41  & 11.20  & 9.92  & 0.7  & 4.2\\
107   & 11 12 24.8  & -76 37 06  &  \object{T51}, \object{Sz41}, \object{CHX20}  &  11.60  &  10.22  & 9.15  & 7.90  & 1.0  & 0.8\\
108   & 11 12 28.1  & -76 44 22  &  \object{T52}, \object{CV Cha}  &  10.96  &  9.53  & 8.22  & 6.75  & 2.2  & 1.2\\
109  & 11 12 28.2  & -76 25 29  &  \object{CHXR53}  &  14.72  &  12.19  & 10.81  & 9.86  & 0.5  & 1.7\\
110   & 11 12 31.4  & -76 44 24  &  \object{T53}, \object{CW Cha}  &  14.45  &  13.50  & 11.22  & 9.22  & 1.5  & 4.4\\ 
111  & 11 12 42.6  & -76 58 40  &  \object{CHX21a}  &  &  11.56  &  10.39  &  9.35  &  2.1  & 1.1\\
112   & 11 12 43.1  & -77 22 23  &  \object{T54}, \object{CHX22}  &  11.16  &  9.64  & 8.58  & 7.71  & 2.3  & 0.8\\
113  & 11 12 43.4  & -76 37 05  &  \object{CHX20E}  &  &  10.95  &  10.01  &  9.08  &  0.6  & 0.4\\
114   & 11 12 49.1  & -76 47 07  &  \object{Hn17}  &  &  13.66  & 12.07  & 11.01  & 2.1  & 2.4\\
115  & 11 13 20.6  & -77 01 04  &  \object{CHXR57}  &  14.63  &  12.27  & 10.87  & 9.79  & 0.8  & 1.8\\
116   & 11 13 24.9  & -76 29 23  &  \object{Hn18}  &  &  13.51  & 11.78  & 10.64  & 0.2  & 2.7\\
\hline
\end{tabular}
\end{table*}

\begin{table*}
{\bf Table \ref{knownYSO}.} (continued)\\
\begin{tabular}{lcclrrrrrr}
\hline
Nb & R.A. (J2000) & Dec. (J2000) & Name\footnotemark[1] & $V$ & $I$ &$J$ &$K_{\rm s}$ & 
Av$_{cl}$ & Av$_*$\\
(1)&(2)&(3) &(4) &(5) &(6) &(7) &(8) &(9) & (10)\\
\hline
117  & 11 13 27.8  & -76 34 17  &  \object{CHXR59}  &  14.35  &  12.09  & 10.54  & 9.55  & 0.1  & 2.4\\
118   & 11 13 30.1  & -76 29 01  &  \object{Hn19}  &  &  13.40  & 11.53  & 10.45  & 0.1  & 3.0\\
119   & 11 13 34.0  & -76 35 37  &  \object{T55}, \object{Sz44}  &  &  13.48  & 11.69  & 10.71  & 0.1  & 3.0\\
120   & 11 14 16.1  & -76 27 37  &  \object{Hn20}  &  &  13.21  & 11.29  & 10.00  & 0.0  & 3.4\\
121   & 11 14 24.9  & -77 33 07  &  \object{Hn21W}  &  &  14.36  & 12.04  & 10.57  & 1.6  & 3.3\\
122   & 11 14 26.6  & -77 33 05  &  \object{Hn21E}  &  &  16.19  & 12.91  & 11.69  & 1.6  & 7.3\\
123  & 11 14 50.8  & -77 33 39  &  \object{B53}  &  &  11.89  &  10.36  &  9.41  &  1.4  & 2.3\\
124  & 11 16 13.4  & -77 14 07  &  \object{CHXR65}  &  13.33  &  11.88  & 10.87  & 10.16  & 0.0  & 0.6\\
125   & 11 17 37.4  & -77 04 38  &  \object{T56}, \object{Sz45}  &  13.50  &  11.78  & 10.24  & 9.27  & 0.2  & 2.2\\
126  & 11 18 20.6  & -76 21 58  &  \object{CHXR68}  &  13.37  &  10.96  & 9.69  & 8.72  & 0.0  & 1.4\\
\hline
\end{tabular}

\begin{description}
\item[$^1$] T{\tt \#} from Schwartz \cite*{Sch91} ; Sz{\tt \#} from Schwartz
	\cite*{Sch77} ; C{\tt \#} from Hyland et al.
	\cite*{HJM82}, Jones et al. \cite*{JHH+85} ; CHX{\tt \#} from
	Feigelson \& Kriss \cite*{FK89} ; CHXR{\tt \#} from Feigelson et al.
	\cite*{FCMG93}, Lawson et al. \cite*{LFH96} ; Baud{\tt \#} from Baud et 
	al. \cite*{BBW+84} ; Hn{\tt \#} from Hartigan \cite*{Har93} ; 
	B{\tt \#} from Alcal\'a PhD thesis, referenced in Lawson et al.
	\cite*{LFH96} ; Ced110 IRS{\tt \#} from Prusti et al. \cite*{PCW+91}
\item[$^2$] Out of field. 
\item[$^3$] Not detected in $J$ nor in $K_{\rm s}$, $I$ not available.
\item[$^4$] $I$ not available.
\item[$^5$] $I$ not available.
\item[$^6$] Not detected in $K_{\rm s}$.
\item[$^7$] Not detected by DENIS. Previous measurement \cite{PWW92}
		have found $J=13.60$ and $K=12.77$. Since the detection limit is 
16
		in $J$ this object should have important luminosity variation.
\item[$^8$] Not detected by DENIS. Previous measurement \cite{HJM82} have 
		found $J=13.84$ and $K=10.66$. This object is in the line of 
sight of
		the Infrared Nebula (IRN).
\item[$^9$] Not Detected. Measured by Hartigan \cite*{Har93}, $J=18.10$
			$K=14.66$.
\item[$^{10}$] Not Detected. Measured by Hartigan \cite*{Har93}, $J=16.34$
		$K=14.58$.
\end{description}
%%%%%%%%%%%%%%%%%%%%%%%%%%%%%%%%%%%%%%%%%%%%%%%%%%%%%%%%%%%%%%%%%%%%%%%%%%%%%%%%
%%%%%%% Fin de la table 1 %%%%%%%%%%%%%%%%%%%%%%%%%%%%%%%%%%%%%%%%%%%%%%%%%%%%%%
%%%%%%%%%%%%%%%%%%%%%%%%%%%%%%%%%%%%%%%%%%%%%%%%%%%%%%%%%%%%%%%%%%%%%%%%%%%%%%%%
\end{table*}

\begin{table*}
	\caption[]{Position and DENIS colours of new YSO candidates}
	\label{newYSO}
%%%%%%%%%%%%%%%%%%%%%%%%%%%%%%%%%%%%%%%%%%%%%%%%%%%%%%%%%%%%%%%%%%%%%%%%%%%%%%%%
%%%%%%% Debut de la table 2 (candidate TTS) %%%%%%%%%%%%%%%%%%%%%%%%%%%%%%%%%%%%
%%%%%%%%%%%%%%%%%%%%%%%%%%%%%%%%%%%%%%%%%%%%%%%%%%%%%%%%%%%%%%%%%%%%%%%%%%%%%%%%
\begin{tabular}{lcccrrrrr}
\hline
Nb & Id. name & R.A. (J2000) & Dec. (J2000) & $I$ &$J$ &$K_{\rm s}$ & Av$_{cl}$ & 
Av$_*$ \\
(1)&(2)&(3) &(4) &(5) &(6) &(7) &(8) &(9)\\
\hline
1   &  DENIS-P J1058.4-7826  &  10 58 22.9  &  -78 26 48  &  11.09  &  9.08  &  7.50  &  0.4  & 3.6\\
2   &  DENIS-P J1059.2-7826  &  10 59 12.3  &  -78 26 39  &  10.65  &  8.93  &  7.47  &  0.1  & 2.8\\
3   &  DENIS-P J1101.1-7730  &  11 01 03.2  &  -77 30 35  &   &  15.85  &  12.86  &  2.9  & \\ 
4   &  DENIS-P J1101.5-7750  &  11 01 29.5  &  -77 50 41  &  17.39  &  16.10  &  13.14  &  4.2  & 1.5\\
5   &  DENIS-P J1101.5-7742  &  11 01 32.1  &  -77 42 10  &  13.57  &  10.73  &  8.68  &  3.8  & 6.1\\
6   &  DENIS-P J1102.4-7753  &  11 02 23.9  &  -77 53 23  &  16.78  &  13.14  &  10.88  &  5.0  & 8.5\\
7   &  DENIS-P J1102.4-7753  &  11 02 25.3  &  -77 53 15  &   &  15.86  &  12.95  &  5.0  & \\ 
8   &  DENIS-P J1102.8-7738  &  11 02 47.1  &  -77 38 09  &   &  13.75  &  10.50  &  7.9  & \\ 
9$^\star$   &  DENIS-P J1103.2-7736  &  11 03 11.5  &  -77 36 36  &   &  13.40  &  10.01  &  8.4  & \\ 
10   &  DENIS-P J1103.5-7749  &  11 03 31.6  &  -77 49 02  &   &   &  12.77  &  9.4  & \\  
11   &  DENIS-P J1103.8-7747  &  11 03 45.7  &  -77 47 04  &   &   &  12.83  &  8.6  & \\  
12   &  DENIS-P J1104.2-7750  &  11 04 11.1  &  -77 50 13  &  18.48  &  13.88  &  10.64  &  9.0  & 11.4\\
13   &  DENIS-P J1105.0-7721  &  11 05 00.6  &  -77 21 47  &  16.26  &  14.28  &  12.84  &  3.8  & 3.5\\
14   &  DENIS-P J1105.3-7757  &  11 05 19.2  &  -77 57 39  &  18.13  &  13.74  &  10.34  &  6.8  & 10.8\\
15   &  DENIS-P J1105.3-7723  &  11 05 20.9  &  -77 23 02  &  16.85  &  14.00  &  12.16  &  4.7  & 6.2\\
16   &  DENIS-P J1105.4-7607  &  11 05 24.0  &  -76 07 44  &  11.99  &  10.00  &  8.41  &  0.4  & 3.6\\
17   &  DENIS-P J1105.9-7738  &  11 05 54.4  &  -77 38 43  &  15.52  &  11.52  &  8.90  &  6.3  & 9.6\\
18   &  DENIS-P J1105.9-7735  &  11 05 55.1  &  -77 35 13  &  17.98  &  13.19  &  9.75  &  6.4  & 12.0\\
19   &  DENIS-P J1106.3-7737  &  11 06 15.9  &  -77 37 51  &  16.55  &  12.72  &  10.08  &  6.5  & 9.1\\
20   &  DENIS-P J1106.3-7735  &  11 06 18.9  &  -77 35 18  &   &  16.21  &  12.93  &  6.7  & \\ 
21$^\star$   &  DENIS-P J1107.2-7718  &  11 07 09.6  &  -77 18 47  &   &  15.12  &  11.32  &  9.9  & \\ 
22   &  DENIS-P J1107.2-7646  &  11 07 10.1  &  -76 46 22  &  13.95  &  11.33  &  9.35  &  2.0  & 5.5\\
23   &  DENIS-P J1107.3-7723  &  11 07 16.8  &  -77 23 07  &   &   &  11.83  &  7.1  & \\  
24$^\star$   &  DENIS-P J1107.4-7722  &  11 07 21.9  &  -77 22 12  &   &  15.28  &  10.86  &  9.3  & \\ 
25$^\star$   &  DENIS-P J1107.4-7741  &  11 07 24.4  &  -77 41 26  &  12.32  &  9.65  &  7.40  &  5.1  & 5.6\\
26   &  DENIS-P J1107.6-7735  &  11 07 37.2  &  -77 35 17  &  16.31  &  12.45  &  9.90  &  5.9  & 9.2\\
27   &  DENIS-P J1107.6-7733  &  11 07 37.8  &  -77 33 10  &  18.03  &  13.68  &  10.81  &  6.2  & 10.7\\
28   &  DENIS-P J1107.8-7738  &  11 07 45.6  &  -77 38 06  &  17.16  &  11.94  &  8.09  &  5.1  & 13.3\\
29   &  DENIS-P J1107.9-7743  &  11 07 55.7  &  -77 43 56  &  18.35  &  15.69  &  13.54  &  4.0  & 5.6\\
30   &  DENIS-P J1108.0-7737  &  11 07 58.2  &  -77 37 21  &   &  15.83  &  12.78  &  4.5  & \\ 
31   &  DENIS-P J1108.0-7703  &  11 08 00.1  &  -77 03 54  &  17.82  &  16.02  &  11.88  &  1.3  & 3.0\\
32$^\star$   &  DENIS-P J1108.1-7738  &  11 08 03.4  &  -77 38 43  &  14.39  &  11.64  &  8.24  &  4.2  & 5.8\\
33   &  DENIS-P J1108.2-7718  &  11 08 12.0  &  -77 18 54  &   &  14.54  &  10.48  &  10.3  & \\ 
34$^\star$   &  DENIS-P J1108.2-7719  &  11 08 12.9  &  -77 19 13  &   &  13.10  &  8.83  &  10.3  & \\ 
35   &  DENIS-P J1108.9-7743  &  11 08 56.5  &  -77 43 30  &   &  15.10  &  12.40  &  3.8  & \\ 
36   &  DENIS-P J1109.2-7632  &  11 09 11.0  &  -76 32 51  &   &  14.42  &  11.68  &  5.5  & \\ 
37   &  DENIS-P J1109.2-7739  &  11 09 11.5  &  -77 39 06  &  16.14  &  12.70  &  10.29  &  5.1  & 7.9\\
38   &  DENIS-P J1109.4-7736  &  11 09 21.9  &  -77 36 54  &  16.74  &  13.93  &  12.00  &  4.0  & 6.0\\
39   &  DENIS-P J1109.4-7631  &  11 09 23.3  &  -76 31 14  &  18.20  &  14.34  &  11.70  &  5.0  & 9.2\\
40   &  DENIS-P J1109.4-7633  &  11 09 26.4  &  -76 33 34  &  17.77  &  13.14  &  10.11  &  6.3  & 11.5\\
41   &  DENIS-P J1109.5-7633  &  11 09 28.9  &  -76 33 28  &   &   &  12.24  &  6.3  & \\  
42   &  DENIS-P J1109.6-7710  &  11 09 38.1  &  -77 10 41  &  15.28  &  11.49  &  8.97  &  3.4  & 9.0\\
43   &  DENIS-P J1109.7-7633  &  11 09 43.5  &  -76 33 30  &   &  15.71  &  12.09  &  6.3  & \\ 
44   &  DENIS-P J1109.8-7634  &  11 09 48.1  &  -76 34 06  &   &  15.17  &  11.83  &  5.6  & \\ 
45   &  DENIS-P J1109.8-7714  &  11 09 49.2  &  -77 14 38  &   &  15.19  &  12.06  &  4.6  & \\ 
46   &  DENIS-P J1109.9-7717  &  11 09 53.4  &  -77 17 15  &   &  15.16  &  12.20  &  5.9  & \\ 
47   &  DENIS-P J1110.0-7718  &  11 09 57.1  &  -77 18 25  &  16.58  &  12.33  &  9.72  &  7.7  & 10.4\\
48$^\star$   &  DENIS-P J1110.2-7635  &  11 10 11.8  &  -76 35 29  &  13.90  &  10.77  &  8.51  &  4.1  & 7.0\\
49$^\star$   &  DENIS-P J1110.9-7725  &  11 10 54.0  &  -77 25 02  &  17.51  &  13.28  &  10.83  &  3.7  & 10.3\\
50   &  DENIS-P J1111.5-7609  &  11 11 29.5  &  -76 09 29  &  12.78  &  9.88  &  7.99  &  1.3  & 6.3\\
51$^\star$   &  DENIS-P J1111.5-7728  &  11 11 32.3  &  -77 28 11  &  14.90  &  11.63  &  9.43  &  3.8  & 7.4\\
52   &  DENIS-P J1113.0-7616  &  11 12 59.8  &  -76 16 53  &  11.93  &  10.09  &  8.56  &  0.2  & 3.1\\
53   &  DENIS-P J1114.2-7630  &  11 14 13.7  &  -76 30 54  &  16.47  &  15.30  &  12.91  &  0.0  & 1.1\\
54   &  DENIS-P J1114.3-7740  &  11 14 15.0  &  -77 40 57  &  16.63  &  15.04  &  12.34  &  1.3  & 2.4\\
\hline
\end{tabular}
\\
$^\star$ object detected by ISO in the LW2 and LW3 filter (P. Persi, private
communication)
%%%%%%%%%%%%%%%%%%%%%%%%%%%%%%%%%%%%%%%%%%%%%%%%%%%%%%%%%%%%%%%%%%%%%%%%%%%%%%%%
%%%%%%% Fin de la table 2 %%%%%%%%%%%%%%%%%%%%%%%%%%%%%%%%%%%%%%%%%%%%%%%%%%%%%%
%%%%%%%%%%%%%%%%%%%%%%%%%%%%%%%%%%%%%%%%%%%%%%%%%%%%%%%%%%%%%%%%%%%%%%%%%%%%%%%%
\end{table*}

\section{Selection  method}
\label{selection}

\subsection{Description}
Our  selection method  of new YSO candidates is essentially based on the  IR 
colours and magnitudes 
of the objects after de--reddening. However, before attempting to select new YSO 
candidates,
we have first  estimated the  contamination by  background and foreground stars 
and the influence
of the luminosity classes on the star magnitudes. To derive this information,  
we have run the so called {\em Besan\c{c}on} model \cite{RC86} in the
direction of the Cha I cloud. The model has been parametrised to
take into account the DENIS limiting magnitudes and  photometric errors. 
It neglects  the   cloud itself (extinction,  star formation).
Figure \ref{besancon} displays synthetic colour-magnitude diagrams obtained
for the different classes of luminosity using this model.
The brightest stars ($K_{\rm s}<8$) correspond  to highly red giants
($J-K_{\rm s} > 1$), and  the colour dispersion of the faintest stars results 
of the photometric errors. We must keep in mind these two points before using
an infrared excess criterion to identify the YSOs. It is clear that we can 
identify only the objects exhibiting a strong infrared excess, i.e $J-K_{\rm s} > 1$.
The number of objects detected both in $J$ and $K_{\rm s}$ band within the covered
area is $\sim$ 10\,000, while the model yields only some 150 foreground
stars, i.e. less than $2 \%
$, assuming a distance of the cloud of  140 pc. 

A previous investigation based on DENIS star counts in the $J$ band on the
Cha I cloud enabled us  to draw an accurate extinction map \cite{CEC+97} with
a spatial resolution of $2 \arcmin$. The peak of visual absorption  that was
measured  is about 10 magnitudes. This map is used to deredden all the 
stars detected within the area that it encompasses. For those  stars located 
inside  the cloud, this reddening is,  of course, an upper limit.

The  YSO candidates are selected according to their colour
and magnitude properties after this dereddening has been applied. 
Consequently, we introduce a bias in the  selection of the reddest 
objects that is discussed below.

In practice, we  have plotted the dereddened magnitudes in a colour-magnitude
diagram together with the main sequence (Fig. \ref {know_new}) and 
selected the stars which are separated from the main sequence by a distance
corresponding to 8 magnitudes of visual extinction, at least.
This provides 90  stars. Part of them cannot be shifted towards the main 
sequence  just assuming  an even  larger  extinction, and  are likely to 
be intrinsically  very red. This sample, still contains some unreliable sources 
because of  photometric errors and, also, some red giant background stars. 
After eliminating these  objects  which are, basically, the bright and the 
faint ends of the sample, we are left with the  54 good candidates  listed
in Table \ref{newYSO}. All of them have been carefully 
checked afterwards by visual inspection of the DENIS images  to 
avoid possible misleading cross-identifications between the 3 DENIS
channels, or optical artifacts such as ghosts  produced by  nearby bright 
star or bad pixels.

\afterpage{\clearpage}

\subsection{Validity}
Since the exact value of the  extinction suffered by each star cannot be 
accurately determined,
we  have assumed that it is  the total extinction  measured on the line of 
sight and taken it as  an upper limit.
Consequently, the constraints on the star colours depend on the location
of the star with respect to the cloud. The criterion is more strict for stars
in the front edge of the cloud than for stars just behind the cloud. 
In other words, the infrared excess can be hidden by  an 
overestimate of the reddening actually suffered by the stars. 
The  stars that we possibly missed  should be, in average, brighter than
the selected stars because they are less obscured. 
Fortunately, the near--infrared range  is less sensitive to extinction than 
optical
bands. The average  visual extinction in the Cha I cloud is about 4 magnitudes,
i.e. $\la 0.5$ magnitudes of $K_{\rm s}$ extinction. 
So, this  effect does not introduce a significant bias in the luminosity 
function. We can crudely  evaluate the maximum number of missed stars.
Assuming that the  extinction  would be reduced by a factor 2, the total
number of selected star with our infrared excess method would increase by 
a factor 1.7, i.e. $\sim 37$ additional stars.
Most of them would probably be background stars because the extinction is now
underestimated for these stars and thus, a residual infrared excess remains.
Nevertheless, this provides an indication of the maximum number of stars that
we can miss.

\section{Discussion}

\subsection{Nature of the sources}
\label{nature}
\begin{figure}[htb]
	\epsfig{figure=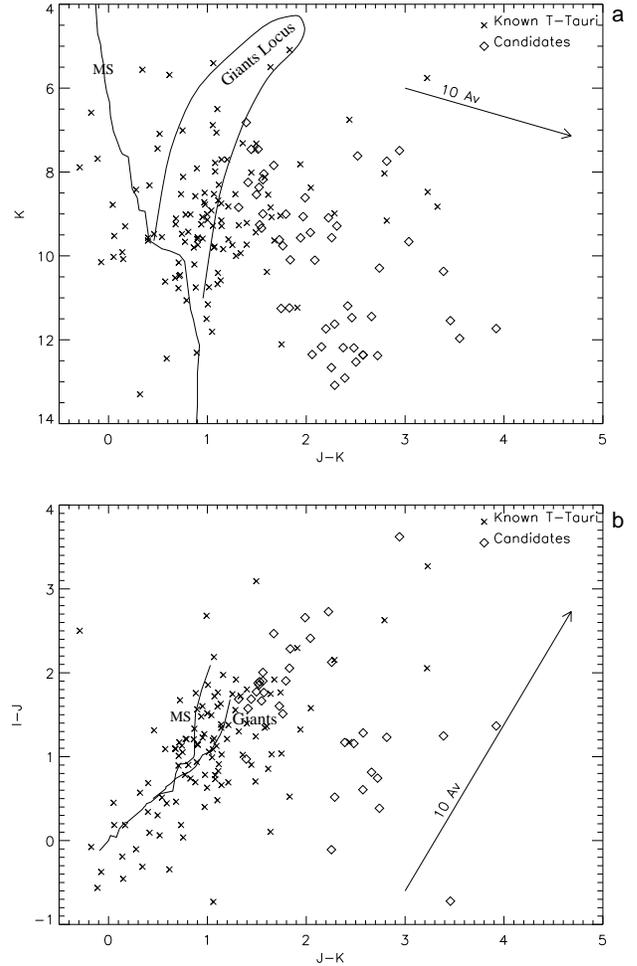,width=8.8cm}
\caption[]{Dereddened diagram for known T Tauri stars and new DENIS
	selected candidates. (a) colour-magnitude diagram for 117 known
	stars and 50 new YSO candidates. (b) colour-colour diagram for 115
	known stars and 34 new YSO candidates. The main sequence (a and b),
	the giants branch (b) and the extinction vector are also plotted}
\label{know_new}
\end{figure}

In the final  sample of 54 candidates, 34 are detected in the 3 channels,
16 only in $J$ and $K_{\rm s}$, and  4 only in $K_{\rm s}$ band (see Table \ref{newYSO}).
Figure \ref{know_new} displays a colour-magnitude and a colour-colour 
diagram for the known T Tauri Stars \cite{FK89,Sch91,PCW+91,GS92,Har93,LFH96}
and the DENIS candidates. The magnitudes given for the known T Tauri Stars 
(TTS) come from the DENIS observations. 

\noindent
The known TTS  consists of Classical and Weak-line TTS \cite{FCMG93}.
The weak-line or naked \cite{FK89} TTS show only  little infrared
excess and they are characterised by a $EW(H_\alpha) < 10$ \AA. It 
corresponds to a later stage of evolution when a large fraction of the 
circumstellar material has been accreted. Hence,  our candidates are
likely to be classical TTS with strong emission lines and massive circumstellar
accretion disks.
We stress the point that our estimation of reddening is an upper limit.
This explains that already known TTS are represented on the left of
the main sequence in the colour-magnitude diagram and in 
the lower left corner of the colour-colour diagram.
%*****
These objects probably lie on the front  side  of the cloud and thus do not
suffer the total extinction that we have derived   on the line of sight. 

The brightest candidates are represented in an area of the diagram where 
already known TTS  are found. Most of the known TTS have a smaller 
value of $J-K_{\rm s}$, but if we try to  extend our sample toward bluer colours, 
we would enter an area corresponding to  the giant and sub-giant stars, where
no separation can  reliably be made on the basis of near-IR colours. 

Moreover, very few known  TTS are as red as the DENIS candidates. An explanation
could be an underestimation of the dereddening, but the colour-colour diagram
indicates that at least 13 objects (detected in the 3 colours) cannot be
shifted toward the main sequence simply by removing an additional reddening 
component.  These candidates are among the reddest objects detected by DENIS.
They were not  previously  detected  because they are too faint and/or 
too far away of the cores.
The cross-identification of DENIS and ROSAT pointed observations in the Cha I
\cite{FCMG93} shows that all X-ray sources detected by ROSAT are brighter than
$K_{\rm s} \simeq 11$ and they do not exhibit strong infrared excess.

\begin{figure}[htb]
	\epsfig{figure=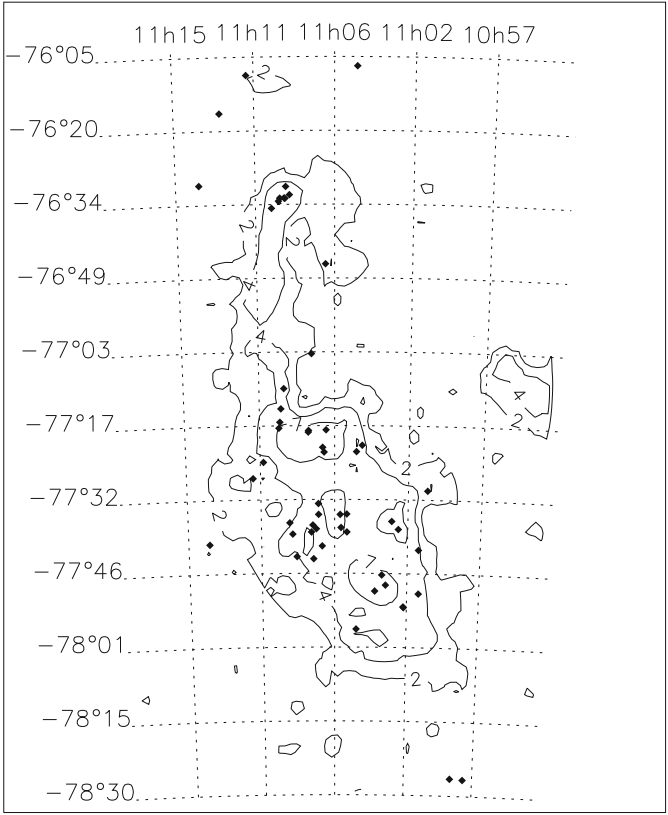,width=8.8cm}
\caption[]{Spatial distribution of YSOs candidates. Extinction isocontour
	at 2,4 and 7 $A_{V}$ are  also plotted}
\label{flag}
\end{figure}

The spatial distribution of the  candidates (Fig. \ref{flag}) is also very
remarkable. These sources have a trend  to be concentrated near the cores of
the cloud which strongly argue in favour of their young nature, since 
background giant stars would be uniformly spread over the whole
field. Another interpretation remains: we could be in presence of small
dark clumps that were not  resolved in  our extinction map. Then, an additional 
reddening 
would have to be taken into account, but, again, 
only a fraction of the candidate locations in colour--colour diagrams can
be explained by normal reddening. In such case, the clumps should be smaller 
than $2 \arcmin$ and  would produce a  visual extinction greater than 15 
magnitudes.
Although this interpretation cannot be  definitely ruled out, we assume that 
these objects 
are more likely to be true  young TTS, since the high infrared excess  probably 
reveals the presence of a large amount of  circumstellar material.

\noindent
Finally, we have examined the interpretation that some of these stars could
be brown dwarfs. On the theoretical evolutionary tracks for low--mass stars,   
young brown dwarfs are actually about 3 magnitudes brighter in $K$ than old 
ones, hence, they could be detected at the distance of Cha I at our 
sensitivity limit. Comer\'on et al. \cite*{CRC+98} have, indeed, reported
the possible discovery of brown dwarfs in the $\rho$ Oph molecular cloud at
a comparable distance of 160 pc. However, the $J-K$ excess of young brown
dwarfs is generally smaller than 1.5 and the $I-J$ is greater than 3.5.
Since our faintest candidates have all $J-K > 2$ and $I-J < 2$, we conclude
that they are unlikely to be brown dwarfs.

\subsection{Evolutionary status of the new candidates}
\label{circum}
In order to evaluate the stage of evolution of these new objects, we have 
attempted to compare their position  in a HR diagram with the evolutionary
tracks modelled by D'Antona \& Mazzitelli \cite*{DM94}. 
Since this  model  does not take into account
the circumstellar contribution, it is  necessary to evaluate the respective 
contributions 
of the stellar and circumstellar components to the emergent flux.  The TTS, 
especially
the classical ones  are surrounded by massive circumstellar accretion disks
which  produces both  extinction and  emission. The extinction is caused by the 
dust
grains, but the emission process is more intricate. Rydgren \& Zak \cite*{RZ87}
have shown that the infrared excess cannot be explained only in terms of
thermal emission of grains in the circumstellar disk. An intrinsic disk
luminosity contribution  is required to account for the observations. They have 
shown
that the intensity of the intrinsic component is directly related to the 
accretion rate and has the same order of magnitude as the thermal emission.
According to Calvet et al. \cite*{CHS97} the near-infrared emission of
protostars is largely dominated by an infalling dusty envelope emission.
Protostars correspond to Class I objects \cite{Lad87,AM94} with massive 
envelopes. TTS are Class II or Class III objects, the essential of their 
circumstellar
matter is in the accretion disk.
Meyer et al. \cite*{MCH97} suggest that the infrared excess of TTS is
essentially due to the disk emission, without any envelope effect. They found
accretion rates in the range from $10^{-8}$ to $10^{-6} M_{\sun}.$yr$^{-1}$ and
inner disk radii from 2 to 6 $R_*$.

Fortunately, the contribution of the disk to the emission is likely to be
negligible  at wavelengths shorter than $\sim 1.6 \ \mu {\rm m}$. Then, the $I$, and,
with probably  less confidence,  the  $J$ fluxes can be considered as mainly
of photospheric origin, although assuming no $I$ excess, Meyer et al.
\cite*{MCH97} have estimated the $J$ excess to be only $10 \%
$ of the photospheric flux for the classical TTS and to 0 for weak--line TTS.
Thus, the colour excess is a good estimator of the extinction suffered by the
star. The use of the $I-J$ colour excess may lead to underestimating the
extinction because of the intrinsic luminosity of the accretion disc, but
probably  not more than 50\%.

In order to convert the evolutionary  tracks in the  HR diagram of D'Antona 
\& Mazzitelli \cite*{DM94}  into similar tracks in 
colour--magnitude diagram we use the Flower's table \cite*{Flo96} which gives
the bolometric correction for the $V$ band versus the effective temperature.
Assuming that the star photosphere emits like a black body at the effective
temperature, we derive $V,I {\rm \ and\ } J$.
%*****
The comparison of the main--sequence that we have constructed in this way with 
main--sequences based upon observations by 
Bessel \& Brett \cite*{BB88} and Johnson \cite*{Joh66}
indicates that the assumption that the photosphere radiates like a black body
is basically  correct for these colours. 
%******
We remark that it is not the case for the G4-M6 spectral range in the $K_{\rm s}$ band.

\begin{figure}[htb]
	\epsfig{figure=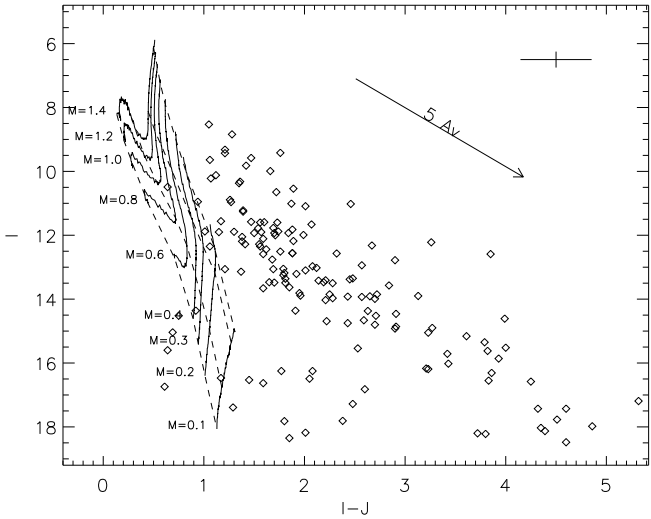,width=8.8cm}
\caption[]{Representation of the associated members of the cloud in a
	colour-magnitude diagram with pre-main-sequence tracks derived from the 
model of
D'Antona \& Mazzitelli (1994). Dashed lines
	are the isochrones at $10^5$, $10^6$, $10^7$ and $10^8$ years}
\label{trackI_IJ}
\end{figure}
\noindent
Previous studies have shown that the associated members of the Cha I 
cloud are essentially in the K7-M5 spectral range \cite{AKJ83,LFH96}. 
Assuming  that all stars are  of spectral type  M0  we can estimate the
extinction they suffer.
Figure \ref{trackI_IJ} shows a $I-J$ versus $I$ diagram where all the TTS
(known and candidates) are represented. 
The representation of the stars is clearly distributed along the
reddening vector. Their position is compatible with a M0 spectral type 
($M \simeq 0.6 M_{\sun}$) except for a part of the faintest sources for which 
$M \la 0.2 M_{\sun}$ seems to be more appropriate. Note that 
22 stars  are missing because they are not detected in $I$. 
The extinction values that we have taken for each individual source are given
in  Tables \ref{knownYSO} and \ref{newYSO}.
The restriction to the M0 spectral type implies uncertainties smaller than
1 magnitude of visual extinction because the tracks for different masses
remain very close to each other.
\begin{figure}[htb]
	\epsfig{figure=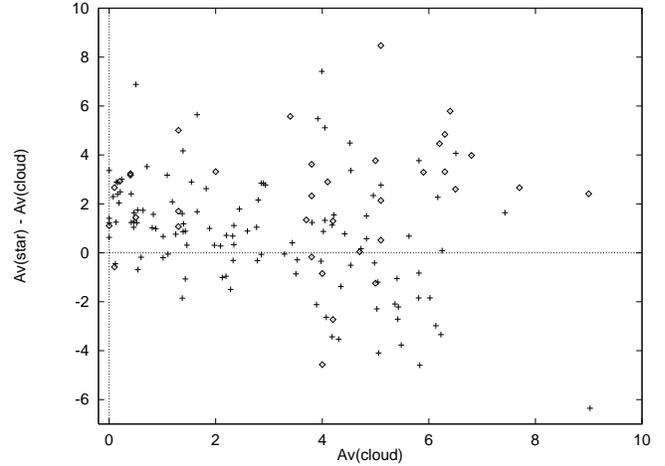,width=8.8cm}
\caption[]{Difference between extinction estimated from star count 
	(cloud extinction) and from $I-J$ colour excess versus cloud
	extinction. Diamonds represent the DENIS candidates and crosses
	the known TTS}
\label{AvAv}
\end{figure}

Without a circumstellar component, we should have $A_{V}(star) \le A_{V}(cloud)$
-- the equality standing for a star just located behind the cloud.
In Fig. \ref{AvAv} we remark that less than $1/3$ of the stars are in
that case. The dispersion of $A_{V}(star)-A_{V}(cloud)$ increases with $A_{V}(cloud)$.
A first explanation could be simply the fact that the location of each star 
within the cloud is not known precisely.
Moreover, the separation between circumstellar and cloud extinction is not
possible. Another explanation could be the evolutionary stage of the star,
the closer a star is to its forming area (i.e. the most obscured regions)
the younger it is likely to be. So, this  dispersion may result from the
effect of the circumstellar disk properties : mass, inclination.
Figure \ref{AvAv} shows that visual extinction for stars is smaller  than 10 
magnitudes which confirms they are likely to be TTS rather than protostars
\cite{LA92}. Protostars have massive envelopes which cause greater extinction
of up to tens of magnitudes. 
Nevertheless, only stars detected in $I$ are represented and, among  the 6 
stars detected only in $K_{\rm s}$, 4 are also detected by ISO (IRS4 and IRS5 in 
Table \ref{knownYSO} ; $nb$ 23 and 41 in Table \ref{newYSO}) in the LW2 
and LW3 filters  centred at $6.75 \ \mu {\rm m}$ and $15.0 \ \mu {\rm m}$, respectively 
(P. Persi, private communication).
These sources are thus likely to be protostars of Class I.
According to Prusti et al \cite*{PCW+91} Ced110 IRS4 and IRS6
are confirmed Class I objects.

\subsection{The Luminosity Function}
\label{KLF}
\begin{figure}[htb]
	\epsfig{figure=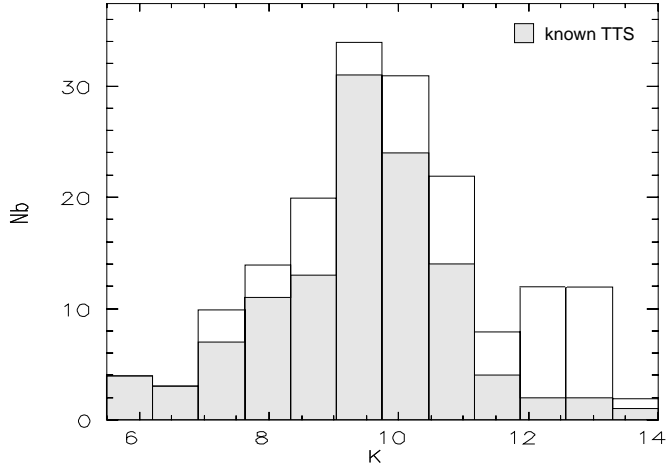,width=8.8cm}
\caption[]{The $K_{\rm s}$ Luminosity Function (KLF) for the associated members
	of the cloud. The previously  known TTS are represented in grey}
\label{histoK}
\end{figure}

\begin{figure}[htb]
	\epsfig{figure=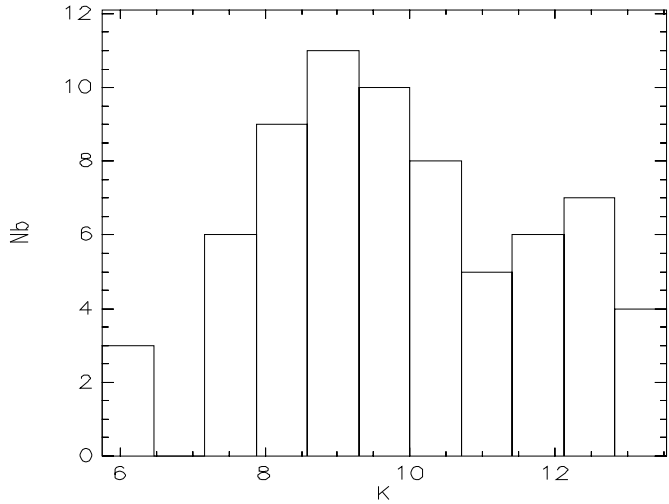,width=8.8cm}
\caption[]{The $K_{\rm s}$ Luminosity Function (KLF) for the DENIS selected
	members of the cloud (new candidates and the 16 reddest known TTS)}
\label{histoK.cand}
\end{figure}

Based on a study of the solar neighbourhood stellar  population,
Miller \& Scalo \cite*{MS79}   have derived the IMF
from the $V$ luminosity function of main-sequence stars, and  concluded that
the IMF can be well approximated by a half-Gaussian distribution. 
In star-forming regions the population consists of YSOs often too faint in $V$
or too highly reddened to be seen  on Schmidt plates.
The $K$ band is the most appropriate to investigate the luminosity function 
in these areas  \cite{LYG93,Meg96,GCNM98}.
Figure \ref{histoK} displays the $K_{\rm s}$ luminosity function (KLF) for the TTS
of the cloud including known sources and our candidates. 
The sample of known stars merge various observations  that may  be  neither  
homogeneous, nor complete, since stars  have been selected according to
various criteria: ${\rm H}\alpha$ emission, near-infrared excess or X-ray emission.
Examination of the KLF (Fig. \ref{histoK.cand}) for all the DENIS selected
objects  shows   2 turnovers, the first one at
$K_{\rm s} \simeq 9.5$ in the KLF is  probably real  while the second  one at $K_{\rm s} 
\simeq 12.5$ corresponds to the DENIS completeness limit. 

\begin{figure}[htb]
	\epsfig{figure=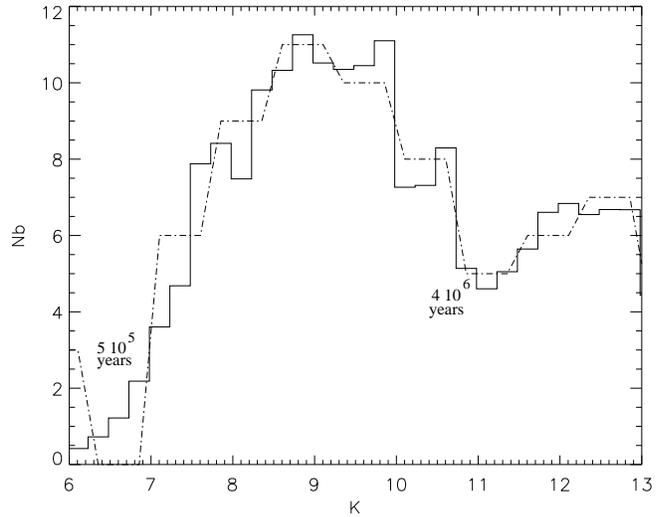,width=8.8cm}
\caption[]{The DENIS KLF (dashed line) and the simulated KLF (solid line)}
\label{KLF_fit}
\end{figure}

The interpretation of the luminosity function requires a model for the star
formation. Theoretical evolutionary tracks \cite{DM94} are used to derive a 
mass-luminosity relation. 
Stellar  $K_{\rm s}$ magnitudes are computed  assuming that they radiate like a  
black-body 
at their  effective temperature and are calibrated using the standard solar
parameters. We compute the mass-luminosity
relation for 44 ages ranging from $10^5$ to $10^8$ years. We obtain similar
results as Zinnecker \& McCaughrean \cite*{ZM91} who have derived the
relation for the ages $3\, 10^5$, $7\, 10^5$, $10^6$, $2\, 10^6$ years.
To estimate when the star formation occurs in the cloud, we use the
half-Gaussian IMF given by Miller \& Scalo \cite*{MS79}.
A Monte-Carlo simulation gives the 44 luminosity functions $\phi_i (K_{\rm s})$ for
each age corresponding to a mass-luminosity relation. 
Since each theoretical KLF, $\phi_i (K_{\rm s})$, corresponds to a given age, 
the observed DENIS KLF can be fitted by a sum of $\phi_i (K_{\rm s})$ with different
weight coefficients $a_i$. 
The solution consists in the resolution of the linear system of equations :
$$
\sum_{i=10^5 yr}^{10^8 yr}{a_i \times \phi_i(K_{\rm s})} = \Phi(K_{\rm s})
$$
\noindent
The singular value decomposition is used to invert the matrix and then,
derive the $a_i$ coefficients. When a coefficient is found to be negative,
the corresponding luminosity function is removed and the process is
iterated with the remaining luminosity functions.
Figure \ref{KLF_fit} shows the simulated KLF
and the DENIS KLF. Because of the small number of stars, the birth rate
function cannot be derived accurately. The Poissonian errors do not allow
the determination of its shape (increasing, constant or decreasing with time).

\noindent
Nevertheless, the presence of two peaks in the KLF allows the estimate of
the age of the sample. At first glance, one could invoke 
another period of star formation  to explain the 
second peak at $K_{\rm s} \simeq 12.5$. This interpretation would, however, 
require stars older than $3\, 10^7$ years which is in contradiction with our 
criterion that selects only stars with massive circumstellar  accretion disks.
Since the accretion rate is known to be about $10^{-7} M_{\sun}.$yr$^{-1}$ 
\cite{MCH97}
our selected stars cannot be older than a few million years and this 
interpretation 
should be ruled out.

To explain this peak, several factors should be taken into account.
The intrinsic dispersion of the magnitudes and the completeness
limit imply a Malmquist bias which causes an overestimation of the number
of faintest stars. This bias is about $5\, 10^{-3}$ magnitude and so, can
be ignored. Examination of the position of the 9 stars fainter than $K_{\rm s}=11$
 and detected in the three colours (9 objects)  in a colour-colour
diagram  shows that their extreme colour cannot be interpreted in terms of 
reddening, hence high extinction clumps on their line of sight cannot be 
invoked.

\noindent
Among the 11 stars which are not detected
in $I$, 3 ($nb$ 36, 43 and 44)  lie in the direction of the molecular
outflow  observed in CO lines by Mattila et al. \cite*{MLT89} in a  22\arcsec\ 
beam size. 
In this direction, they estimate a  visual extinction of 17 
magnitudes.
Since our estimate of the extinction in this area is only 6 magnitudes with
a 2\arcmin\ resolution map, these stars can be background objects.

The 6 stars detected only in $K_{\rm s}$ also contribute to the second peak. 
Some of them are known protostars (IRS4 and IRS6 in Table \ref{knownYSO}) and
others are likely to be so (see above).
Their faintness results from the effect of a massive circumstellar envelope. 
We remark that the star responsible for  the outflow cited above could be
the star $nb$ 41, detected only in $K_{\rm s}$, rather than the known source
T42 = sz32 (Persi P., private communication).
Finally, possible unresolved binaries would lead to  misleading colours that
cannot be corrected. So, this peak is not fully understood and further deeper
observations in $K$ are requested to reach a better completeness limit.\\

The first peak at $K_{\rm s} \simeq 9$ requires a period of star formation
that would extend from  $4\, 10^5$ to $3\, 10^6$ years.
Besides, the $K_{\rm s}$ excess modifies the luminosity function with a mean
of 1.5 magnitude with respect to the spectral energy distribution of
a main--sequence star. The mean extinction suffered by stars is about 5
magnitudes of visual extinction (Fig. \ref{AvAv}), i.e. 0.5 magnitude of $K_{\rm s}$
extinction.
The combination of these two effects shifts the KLF of 1 magnitude toward the
fainter magnitudes. That led us to change our age estimation from
$4\, 10^5$\,--\,$3\, 10^6$ to $5\, 10^5$\,--\,$4\, 10^6$ years.
We stress the point that it just means that we do not
detect older stars with our infrared excess criterion. Older stars have
lost their disk, or, the accretion rate has become smaller and then, they 
no longer exhibit an infrared excess.
The apparent drop of the KLF down
to $K \simeq 11$ probably results from a vanishing of the circumstellar disk
rather than  from a  real decrease of the star formation rate.
The maximum life time of a circumstellar disk is then estimated to 
$4\, 10^6$ years for low-mass stars which corresponds to the oldest
Classical TTS.

\section{Conclusion}

The investigation of the DENIS data obtained in the Cha I cloud allowed 
the identification of 54 new faint YSOs candidates. They are selected on the
basis of a large infrared excess and their concentration near to the cores of
the cloud. 
Because of their IR excess they are likely to be surrounded by a massive
circumstellar disk for which the visual extinction can reach 10 magnitudes
(Fig. \ref{AvAv}) and thus are probably Classical TTS.
Comparison of the different magnitudes in the three DENIS channels shows that
the $K_{\rm s}$ excess is $\sim 1.5$ magnitudes, in agreement with Meyer et al.
\cite*{MCH97} who estimate from spectroscopic observations a $K$ excess
smaller than 1.6 magnitudes.
Investigations on the $K_{\rm s}$ Luminosity Function down to $K_{\rm s} = 13$ shows
important results concerning the age of our sample. First of all, 
youngest stars in the Cha I cloud are about $5\, 10^5$ years, except
for few protostars.
Moreover, the decrease of the KLF for $K_{\rm s} > 10$ indicates that stars older
than $4\, 10^6$ years have lost their circumstellar disk. 
The maximum life time of active disk is estimated to $4\, 10^6$ years, after
this period, the accretion becomes too small ($< 10^{-8} M_{\sun}$.yr$^{-1}$) to
produce an infrared excess.
Spectroscopic follow-up of this new sample is underway to confirm definitely
the nature of the newly discovered objects.

\begin{acknowledgements}
The DENIS team is warmly thanked for making this work possible and in
particular the operations team at La Silla.
The DENIS project is  supported by  the
European Southern Observatory, in France by the {\it Institut National des 
Sciences de l'Univers}, the Education Ministry and the 
{\it Centre National de la Recherche Scientifique}, in Germany by  the State of 
Baden-Wurttemberg, 
in Spain by the DGICYT, in Italy by the Consiglio Nazionale delle Ricerche,
in Austria by the Science Fund and Federal Ministry of
Science, Transport and the Arts, in Brazil by the Foundation for the development
of Scientific Research of the State of S\~ao Paulo (FAPESP), and formerly by 
the {\it SCIENCE}  and the  {\it Human Capital and Mobility} 
plans of the European Commission.
\end{acknowledgements}

%\bibliographystyle{astron}
%\bibliography{mnemonic,biblio,preprint}

\end{document}